\documentclass{article}%[a4paper,10pt]
\usepackage[dvips]{graphicx}
\usepackage{amssymb}
\usepackage{amsmath}
\usepackage{amscd}
\usepackage{mathrsfs}
\usepackage{subfigure}
\usepackage{latexsym}
\usepackage{makeidx}
\date{}
\author{Tina A.C. Maiolo\footnote{Dipartimento di Fisica dell'Universit\`a del Salento and Sezione INFN di Lecce, 73100 Lecce, Italy; e--mail: tina.maiolo@le.infn.it}, Luigi Martina\footnote{Dipartimento di Fisica dell'Universit\`a del Salento and Sezione INFN di Lecce, 73100 Lecce, Italy; e--mail: luigi.martina@le.infn.it}, Giulio Soliani\footnote{Dipartimento di Fisica dell'Universit\`a del Salento and Sezione INFN di Lecce, 73100 Lecce, Italy; e--mail: giulio.soliani@le.infn.it}}
\title{\textbf{A critical theory of quantum entanglement for the Hydrogen molecule}}
\begin{document}
\maketitle
\begin{abstract}
\noindent
In this paper we investigate some entanglement properties for the Hydrogen molecule considered as a two interacting spin $\frac{1}{2}$ (qubit) model. The entanglement related to the $H_{2}$ molecule is evaluated both using the von Neumann entropy and the Concurrence and it is compared with the corresponding quantities for the two interacting spin system. Many aspects of these functions are examinated employing in part analytical and, essentially, numerical techniques. We have compared analogous results obtained by Huang and Kais a few years ago.
In this respect, some possible controversial situations are presented and discussed.
 
\end{abstract}

\section{Introduction and the model \label{primo}}
Entanglement is a physical observable measured by the von Neumann entropy or, alternatively, by the Concurrence of the system under consideration.

The concept of entanglement gives a physical meaning to the electron correlation energy in structures of interacting electrons. The electron correlation is not directly observable, since it is defined as the difference between the exact ground state energy of the many electrons Schr\"odinger equation and the Hartree--Fock energy.

In this paper we discuss the Hamiltonian which describes the Hydrogen molecule regarded as a two interacting spin $\frac{1}{2}$ (qubit) model.

In \cite{hua05} it was argued that the entanglement (a quantum observable) can be used in analyzing the so--called \emph{correlation energy} which is not directly observable. From our point of view, the Hydrogen molecule is dealt with a bipartite system governed by the Hamiltonian
\begin{equation}
\label{2Ising}
\mathcal H_{H_2}=-\frac{J}{2}(1+g)\sigma_{1}\otimes\sigma_{1}-\frac{{J}}{2}(1-g)\sigma_{2}\otimes\sigma_{2}-\mathcal{B}(\sigma_{3}\otimes\sigma_{3}+\sigma_{0}\otimes\sigma_{3}),
\end{equation}
where $\sigma_{i}$ stand for the Pauli matrices ($\sigma_{0}=I$).
 Actually, this model was considered in \cite{hua05} in order to illustrate their method. However, here we will make some interpretative changes. Indeed, from our point of view, the states of an isolated atom are strongly reduced to a system with two energy levels related to the intensity of the magnetic field $\mathcal{B}$. Relatively to this scale, the exchange interaction constant $J$ is usually smaller than $\mathcal{B}$, in order to represent the residual interatomic interactions. From the point of view of quantum chemistry, one may interpret the discrete spectrum as provided by the Hartree--Fock calculations, while the interaction coupling $J$  models the residual multielectronic effects, not taken into account by the mean field approximation.

 For simplicity we limit ourselves to the ferromagnetic phase with $J>0$. The parameter $g$, such that $0\leq g\leq 1$, describes the degree of anisotropy corresponding for $g=0$ to the completely isotropic XY spin model. Conversely, $g=1$ provides the anisotropic XY spin model, the so-called \emph{Ising model}.

We notice that when the atoms are far apart, their interaction is quite weak. This corresponds to a vanishing value of $J$. In this situation the state of the system is completely factorized in the product state of the ground states of the indipendent spins. The corresponding total energy, in unit of $\mathcal B$, is just the sum of the two fundamental levels, ${\cal{E}}_{0}=-2$, which we may consider as the Hartree-Fock approximated fundamental level in molecular structure calculations.

When $J\neq 0$, the fundamental energy eigenvalue is $\cal{E}$$=-\sqrt{4+g^{2}\lambda^{2}}$ in  Region \textbf{I} defined by $0< \lambda \leq \frac{2}{\sqrt{1-g^{2}}}$, otherwise $\cal{E} =-\lambda$ ($\lambda$ means the coupling constant) in Region \textbf{II}, which is the complement of \textbf{I} which respect to positive real axis. The corresponding (non normalized) eigenstates are $|\Psi_{\textbf{I}}\rangle=\Big\{\frac{\sqrt{g^{2}\lambda^{2}+4}+2}{g\lambda}, 0, 0, 1\Big\}$ and $|\Psi_{\textbf{II}}\rangle=\Big\{0, 1, 1, 0\Big\}$, respectively. \emph{In both cases the state is entangled}.

  Since we are dealing with pure states, the von Neumann entropy \cite{nie00} \begin{equation}
 S_{vN} = - {\texttt{Tr}}\Big( \rho_1 log_{2} \rho_1 \Big) 
\end{equation}
is chosen to be  a measurement of the entanglement, where $\rho_1$ is the 1-particle reduced density matrix.  However, for general mixed states other entanglement estimators (for instance, the Concurrence \cite{run03})  have to be used.  In the considered case, one has
\begin{displaymath}
S_{vN, {\bf{I}}}  = - \Big[  \-g^2 \sqrt{g^2 \lambda ^2+4} \log \Big(\frac{1}{2}-\frac{1}{\sqrt{g^2 \lambda^2+4}}\Big) \lambda ^2 
\end{displaymath}
\begin{displaymath}
   -\Big(g^2 \Big(\sqrt{g^2 \lambda ^2+4}+4\Big) \lambda
   ^2+8 \Big(\sqrt{g^2 \lambda ^2+4}+2\Big)\Big) \log
   \Big(\frac{1}{2}+\frac{1}{\sqrt{g^2 \lambda ^2+4}}\Big)\Big]  
\end{displaymath}
\begin{equation}\label{SvNreg1}
\frac{1}{\sqrt{g^2 \lambda ^2+4} \left(g^2 \lambda ^2+2 \sqrt{g^2 \lambda ^2+4}+4\right) \log (4)}
\end{equation}
\begin{equation}\label{SvNreg2}
 S_{vN, {\bf II}}  =  1 .  
\end{equation}
Scrutinizing Eq. (\ref{SvNreg1}) and Eq. (\ref{SvNreg2}) it emerges that the entropy is an increasing  function of the coupling
 constant $\lambda$ in Region $\textbf {I}$, but the state is
 maximally entangled in Region $\textbf {II}$ independently from the anisotropy parameter $g$. One sees that, as it arises graphycally, for $g=1$ the entanglement is a monotonic increasing function of the interaction coupling $\lambda$. Moreover for weak ($< 1$) coupling values it
   is always less than the 30\%. Of course, for large coupling
   constants the entropy approaches 1, meaning that \emph{all levels are equiprobably visited by the considered spin}.

Limiting all further considerations to the case of weak interaction, we observe that at the boundary point $\lambda_{b}  =\frac{2}{\sqrt{1-g^2}}$  a
  discontinuity occurs, signaling a crossing of the lowest eigenvalues and, in a more general context, a quantum phase transition \cite{sac01}.

As it was pointed out in \cite{ose03}, for quantifying the entanglement we can resort to the reduced density matrix. Furthermore, in \cite{woo98}, Wootters has shown that for a pair of binary qubits one can use the concept of Concurrence $\mathcal C$ to measure the entanglement.

The Concurrence reads
\begin{equation}
\mathcal C (\rho)=max(0, \nu_{1}-\nu_{2}-\nu_{3}-\nu_{4}),
\end{equation}
where the $\nu_{i}$'s are the eigenvalues of the Hermitian matrix 
\begin{displaymath}
R=\Big(\sqrt{\rho} \tilde{\rho}\sqrt{\rho}\Big)^{\frac{1}{2}},
\end{displaymath}
where $\tilde{\rho}=(\sigma_{y}\otimes\sigma_{y})\rho^{\ast}(\sigma_{y}\otimes\sigma_{y})$, $\rho^{\ast}$ being the complex conjugate of $\rho$ taken in the standard basis \cite{woo98}.

Some interesting results on the simple model (\ref{2Ising}) of the Hydrogen molecule can be achieved by realizing a comparative study of the von Neumann entropy and the Concurrence.

To this aim, we compute the Concurrence $\mathcal C_{\textbf {I}}$ and $\mathcal C_{\textbf {II}}$, i. e.
\begin{equation}
{\mathcal C}_{ {\bf I}} = g \lambda  \sqrt{\frac{1}{g^2 \lambda ^2+4}},
  \qquad {\mathcal C}_{ {\bf II}} = 1. 
\end{equation}
where \textbf{I} and \textbf{II} refer to Regions \textbf{I} and \textbf{II}, where $0\leq \lambda \leq \frac{2}{1-g^{2}}$, and ${\cal E}=-\lambda$, respectively. \\

\begin{figure}[!h]
\begin{center}
\includegraphics[angle=0, width=0.55\textwidth]{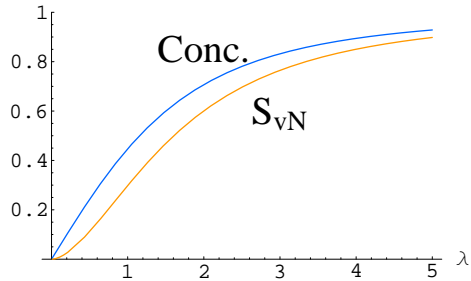}
\caption{Comparison between the Concurrence  and the von Neumann
entropy for the two spins system as a function of the coupling
constant $\lambda$ for $g =1$.}
\label{confronto}
\end{center}
\end{figure}

In Figure \ref{confronto} a comparison between the Concurrence and the von Neumann entropy for two spins system as a function of the coupling $\lambda$ for $g=1$ is presented.\\

Sec. \ref{entcorr} contains a comparison between the entanglement and the correlation energy. In Sec. \ref{qcf} the Configuration Interaction method is introduced to compare entanglement and correlation energy. In Sec. \ref{CI2sIm} some differences between the Configuration Interaction approach and the two spin Ising model are presented. Finally, our main results are summarized in Sec. \ref{fine}.
\section{A comparison between the entanglement and the correlation energy \label{entcorr}}

Now we look for  a  comparison between the entanglement with the  energy
correlation, which as we have already recalled, it is  understood as the difference of the
fundamental energy level compared with respect to the
corresponding value at vanishing coupling constant $\lambda$.

For $g=1$ and in unities of $\mathcal{B}$ it is given by 
\begin{equation}
 {\cal E}_{corr} =
|{\cal E}_0|-2 = {\sqrt{4 + {\lambda }^2}} - 2 .\label{correlazio} 
\end{equation}
We observe that the entanglement measure is \emph{always bounded, while
${\cal E}_{corr}$ is a divergent function of $\lambda$. So it does not
make much sense to look for simple relations valid on the entire
$\lambda$-axes}. Consequently, limiting ourselves to weak couplings, for
$0 \leq \lambda \leq 1$,  we minimize the \emph{mean squared deviation}
\begin{equation} 
I\Big [ \alpha \Big ] = \int_0^1 \Delta S_{ \alpha}^2\, d\lambda,
\qquad {\rm  with}\quad  \Delta S_{ \alpha} =
  {\cal E}_{corr} - \alpha\, S_{vN}. \label{deviation} 
\end{equation}
Thus the minimizing parameter $\alpha_{min}$ will be given by
\begin{equation} 
\alpha_{min} = \frac{\int_0^1  {\cal E}_{corr} S_{vN}\, d\lambda }
  {\int_0^1  S_{vN}^2\, d\lambda } \approx - 0.691217. \label{-0/69} 
\end{equation}

\begin{figure}[!h]
 \begin{minipage}[b]{0.5\textwidth}
   \centering
\includegraphics[angle=0, width=1.0\textwidth]{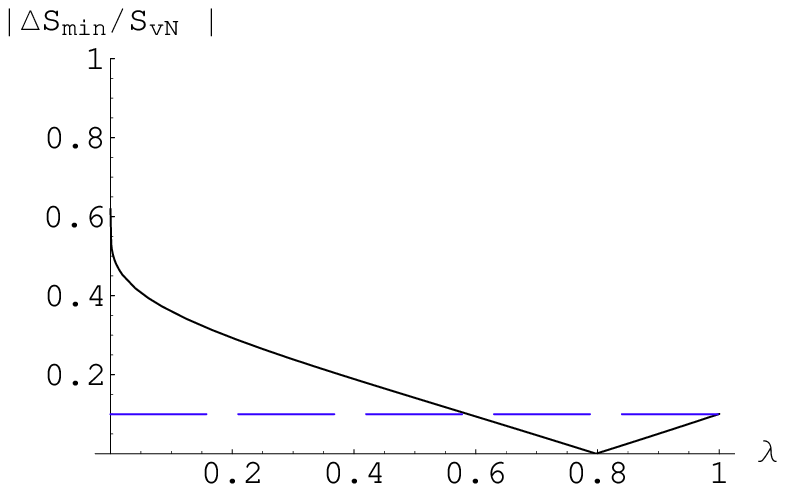}
 \end{minipage}
\begin{minipage}[b]{0.5\textwidth}
   \centering
\includegraphics[angle=0, width=1.0\textwidth]{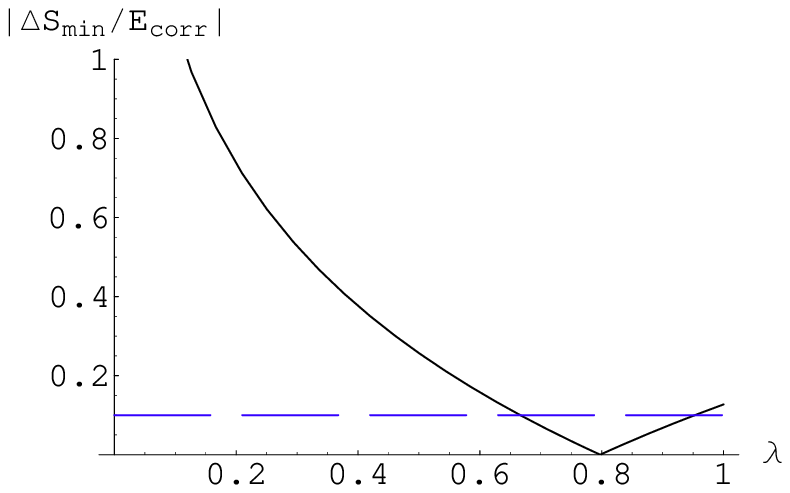}
 \end{minipage}
\caption{The
 relative  quadratic deviation between the  von Neumann entropy and the
correlation energy with respect to the former and the latter, respectively,  at the optimal value $\alpha_{min}$ as a function of the coupling constant $\lambda$ for $g = 1$.}
\label{diple}
\end{figure}

A formula analogous to (\ref{-0/69}) can be obtained by using the Concurrence as a measure of entanglement. In this case, by minimizing the mean squared deviation we have
\begin{equation} 
I_{\mathcal C}\Big[ \alpha' \Big] = \int_0^1 \Delta {\mathcal C}_{ \alpha'}^2\, d\lambda,
\qquad {\rm  with}\quad  \Delta {\mathcal C}_{ \alpha'} =
  {\cal E}_{corr} - \alpha'\, {\mathcal C}. \label{deviationC} 
\end{equation}
Now, in order to estimate the relative deviation of  $ S_{vN}$
with respect to ${\cal E}_{corr}$, let us report $ | \Delta S_{
\alpha_{min}}| /S_{vN} $ and $| \Delta S_{
\alpha_{min}}/ {\cal E}_{corr}| $ as functions of $\lambda$ at the optimal value $\alpha_{min}$. The graphs
of these functions are shown in Figure \ref{diple}.

\begin{figure}[!h]
 \begin{minipage}[b]{0.5\textwidth}
  \centering
\includegraphics[angle=0, width=1.0\textwidth]{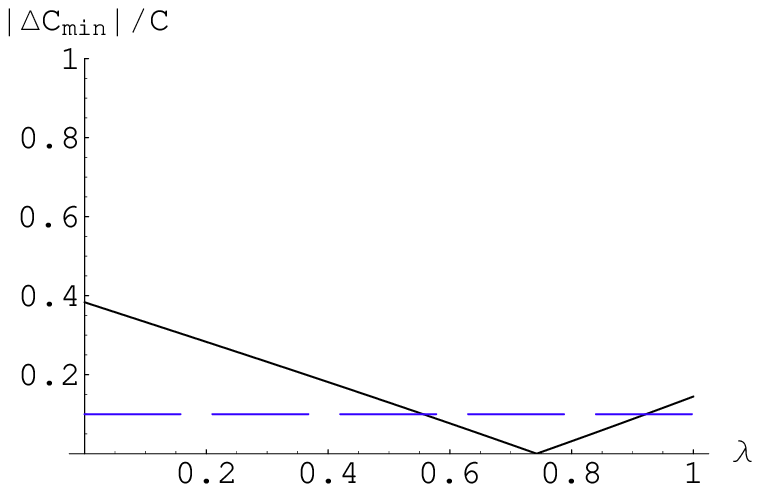}
\end{minipage}
\begin{minipage}[b]{0.5\textwidth}
 \centering
\includegraphics[angle=0, width=1.0\textwidth]{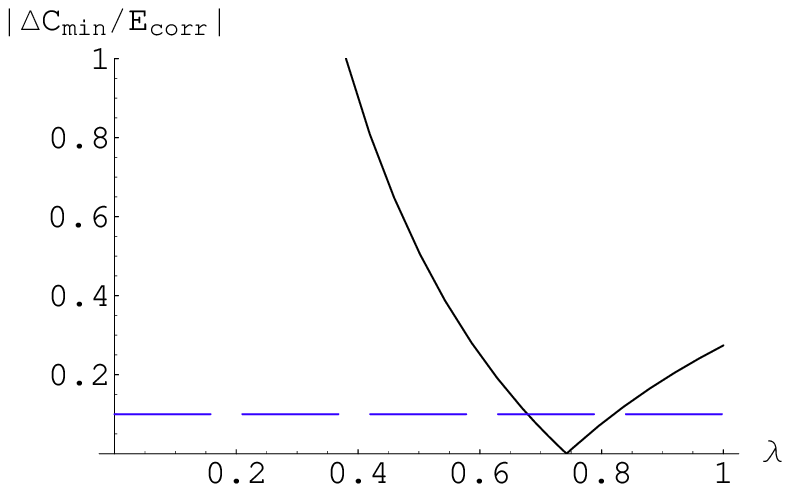}
\end{minipage}
\caption{The
relative  quadratic deviation between the  Concurrence and the
correlation energy with respect to the former and the latter, respectively, at the optimal
value $\alpha'_{min}$ as a function of the coupling constant
$\lambda$ for $g = 1$.}
\label{dConcE}
\end{figure}

In Figure \ref{dConcE}, the relative quadratic deviation between the Concurrence and the correlation energy with respect to the former and the latter, at the optimal values $\alpha ' _{min}$,  is represented.\\

\textbf{Remark 1}\\

From these graphs, one can argue that the agreement between
the two functions $S_{vN} $ and ${\cal E}_{corr}$  is only qualitatively
good, in fact, for very small $\lambda$, it is not good at all. However, in an
intermediate range of values, i. e., $0.6 \leq \lambda \leq 1$ the two
functions are almost proportional within the 10\%. Analogously, the
same is true between energy and Concurrence. Even, the agreement
becomes worst comparing the relative deviation of the Concurrence
with respect to the correlation energy, since the range in which
the relative deviations become smaller than 10\% are narrower.  Then, the
question is whether the above results are i) sufficient to justify the
conjecture advanced in \cite{hua05}, i.e., entanglement can be considered as an estimation of correlation energy; ii) if such a relation has a
more concrete physical meaning, in particular whether the minimizing
parameter $\alpha_{min}$ and the vanishing point of $\Delta
S_{\alpha_{min}}$ does possess any physical meaning (or
$\alpha'_{min}$ and the vanishing point of $\Delta
{\mathcal C}_{\alpha'_{min}}$). Notice that in the case of the comparison for
the Concurrence simpler analytical expressions appear. For instance one finds $\Delta {\mathcal C}_{\alpha'_{min}} = \left(\text{0.383249}
\frac{\lambda}{\sqrt{\lambda ^2+4}}
    -\sqrt{\lambda ^2+4}+2\right)^2$.\\

\textbf{Remark 2}\\

\begin{figure}[!h]
\begin{center}
\includegraphics[angle=0, width=1.0\textwidth]{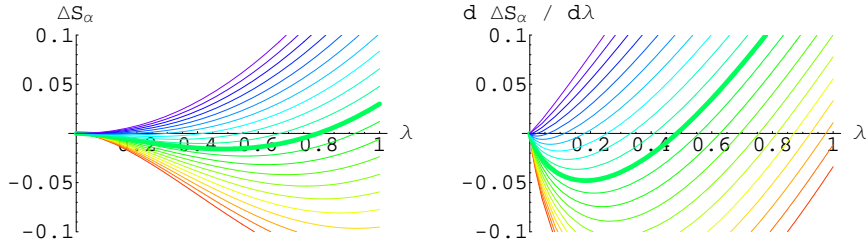}
\caption{The
deviation $\Delta S_{\alpha}$ and its derivative with respect to
$\lambda$ are computed for values of $-1.29 ( \textrm{red}
)\leq \alpha \leq -0.091 ( \textrm{violet} ) $, for steps of
$0.06$. The curve drawn thicker corresponds to $\alpha _{min}$}
\label{aroundminS}
\end{center}
\end{figure}

We note that in an interval of values
around $\alpha_{min}$, the deviation function (\ref{deviation})
possesses a minimum in the interval of interest $0 \leq \lambda
\leq 1$, otherwise  the minimum is achieved at larger value of
$\lambda$, or the function is monotonically increasing (see Figure \ref{aroundminS}).

\begin{figure}[!h]
\begin{center}
\includegraphics[angle=0, width=1.0\textwidth]{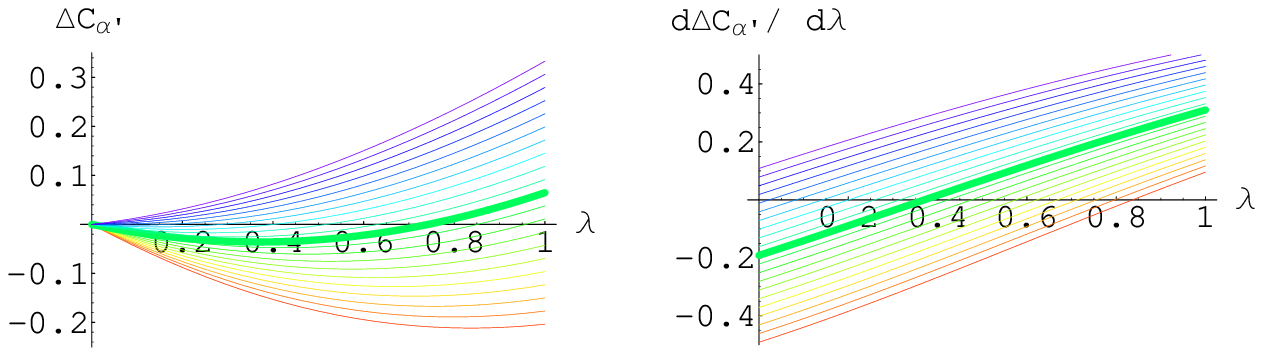}
\caption{The
deviation $\Delta {\mathcal C}_{\alpha '}$ and its derivative with respect to
$\lambda$ are computed for values of $-0.98 ( \textrm{red}
)\leq \alpha ' \leq 0.22 ( \textrm{violet} ) $, for steps of
$0.06$. The curve drawn thicker corresponds to $\alpha' _{min}$}
\label{aroundminC}
\end{center}
\end{figure}

This behavior suggests to consider the function $\Delta
S_{\alpha_{min}}$ as a sort of "free energy" , where
$\alpha_{min}$ mimics the "temperature"  specific
of the system.  If, for some reason, we allow $\lambda$ to
change, then we expect that spontaneously the interaction
coupling adjusts itself to the minimum of $\Delta
S_{\alpha_{min}}$. Similar considerations can be made looking at
the graphs drawn for the function $\Delta {\mathcal C}_{\alpha'_{min}}$ and
its derivative with respect to $\lambda$ (see Figure
\ref{aroundminC}).

The function $\Delta S_{\alpha_{min}}$ or, alternatively, the minimum of $\Delta {\mathcal C}_{\alpha'_{min}}$ can be obtained algebraically. Such a minimum is at the value of the coupling constant ${{\lambda}^{S_{vN}}_{min}\approx {0.485}}$ and ${{\lambda}^{\mathcal C}_{min}\approx {0.371}}$, respectively.

\begin{figure}[!h]
\begin{center}
\includegraphics[angle=0, width=0.55\textwidth]{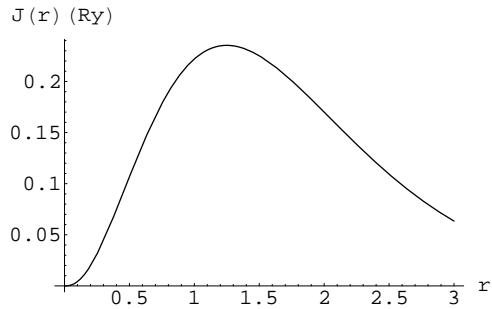}
\caption{The effective
interaction Hydrogen-Hydrogen atom}\label{JrH}
\end{center}
\end{figure}

\begin{figure}[!h]
 \begin{minipage}[b]{0.5\textwidth}
   \centering
\includegraphics[angle=0, width=1.0\textwidth]{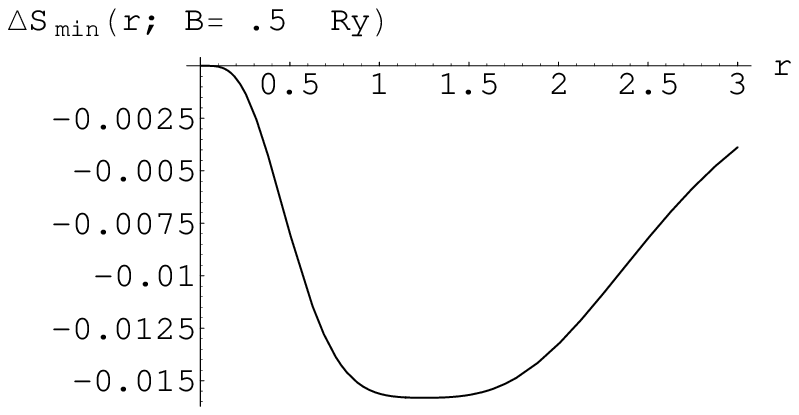}
 \end{minipage}
 \begin{minipage}[b]{0.5\textwidth}
   \centering
\includegraphics[angle=0, width=1.0\textwidth]{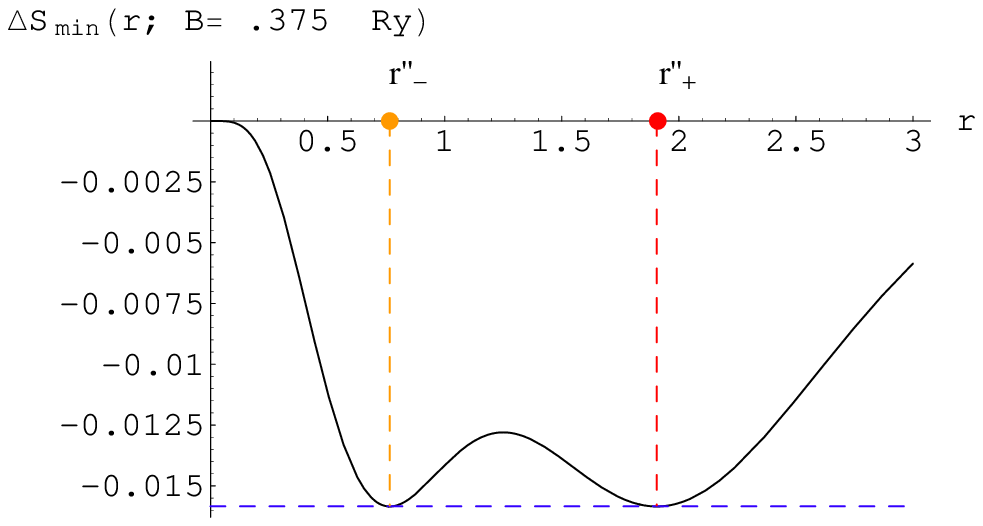}
 \end{minipage}
\caption{The von
Neumann entropy for the 2-spin model for $\mathcal{B} =.5 \;  Ry$ (left panel) and for $\mathcal{B} =.375 \; Ry$ (right panel) and the
position depending interaction given by (\ref{JrHint}).}\label{Svr2}
\end{figure}

The authors in \cite{hua05} studied  numerically
 the von Neumann entropy and the correlation function for
a Hydrogen molecule, using an old result by Herring and Flicker
\cite{her64}, going back  to an oldest idea by Heitler and London \cite{hei27},
 which consists in substituting the  molecular binding with a position
dependent exchange coupling:
 \begin{equation} J(r) \approx  1.641\,r^{\frac{5}{2}}\, e^{- 2\,r} \;{\rm Ry}
 ,\label{JrHint}
\end{equation}  
 where
 $r$ is given in Bohr radius, see Figure \ref{JrH}.
The maximum value taken by this function is   at the point
 $r_{max} = 1.25 $. Assuming $\mathcal{B} = 0.5\; {\rm Ry}$, i.e.
 $\frac{1}{2}$  of the fundamental level of the Hydrogen atom, the maximum
 value  $\lambda'_{max} = J (r_{max})/\mathcal{B} \approx 0.470628
 < \lambda^{S_{vN}}_{min}$, i.e.
  the value of the \emph{effective interaction value is less than the minimum
   for the deviation function} $\Delta S_{\alpha_{min}}$.  Then, the equilibrium balance between
entanglement (as von Neumann entropy)  and correlation energy
predicts a  length of the molecule equal to $r_{max}$ (see the first panel of Figure \ref{Svr2}).
On the other hand,
 if we consider the energy gap $2 \mathcal{B} = 3/4  \;{\rm Ry}$, i.e. the energy step
to the first excited state,  one obtains the new value
$\lambda_{max}''  \approx 0.628 $, which goes beyond  $ {\lambda
}_{min}$, even if it is always less than $1$.   Now, the
deviation function  $\Delta S_{\alpha_{min}}$ has two minima as
seen in the second panel of Figure \ref{Svr2}, one of which is at $ r''_- \approx 0.76
$ , the other one being at $ r''_+ \approx 1.91 $.

These results  should be compared with the experimental equilibrium
length of the Hydrogen molecule, which is $r_{exp} \approx 2.0 $. 

We point out that although the spin--model described by the Hamiltonian (\ref{2Ising}) is characterized by features which are essentially rough, however we are induced to answer positively to the quest for a physical meaning of the deviation function $\Delta S_{\alpha_{min}}$. Indeed, the results elucidated in Figure \ref{Svr2} encourage, on one part, improvement of the computation of $r$ in order to make more accurate the comparison with the experimental value $r_{exp}$.

The first question to answer is whether this draft works also for the
Concurrence. A statement about it is not obvious, since  the von
Neumann entropy is a \emph{nonlinear function} of the Concurrence in the
2-qubits case.

\begin{figure}[!h]
 \begin{minipage}[b]{0.5\textwidth}
   \centering
\includegraphics[angle=0, width=0.70\textwidth]{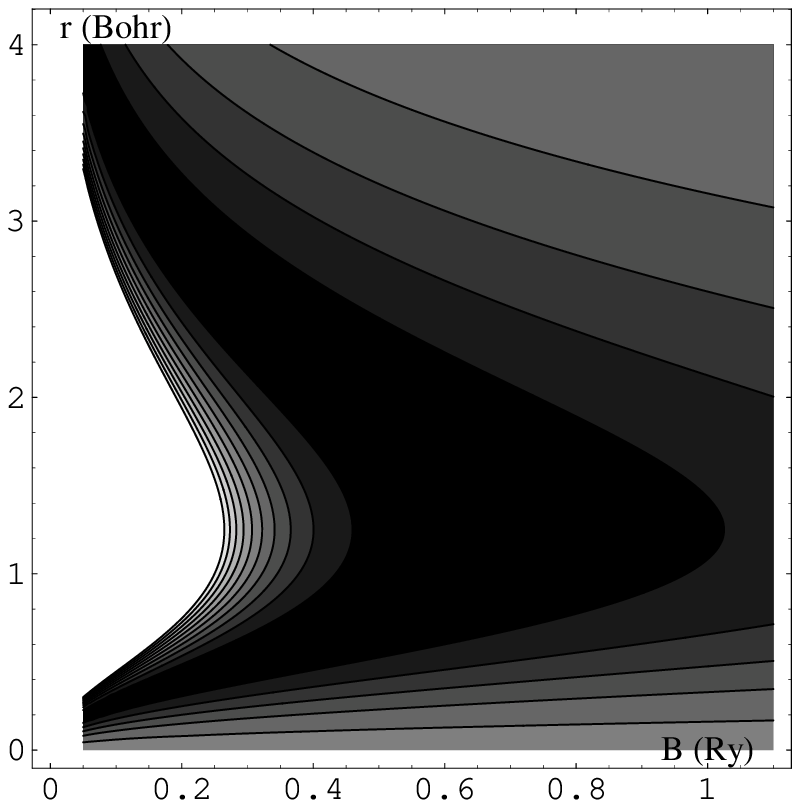}
\label{DCurrHy}
\end{minipage}
 \begin{minipage}[b]{0.5\textwidth}
   \centering
\includegraphics[angle=0, width=0.70\textwidth]{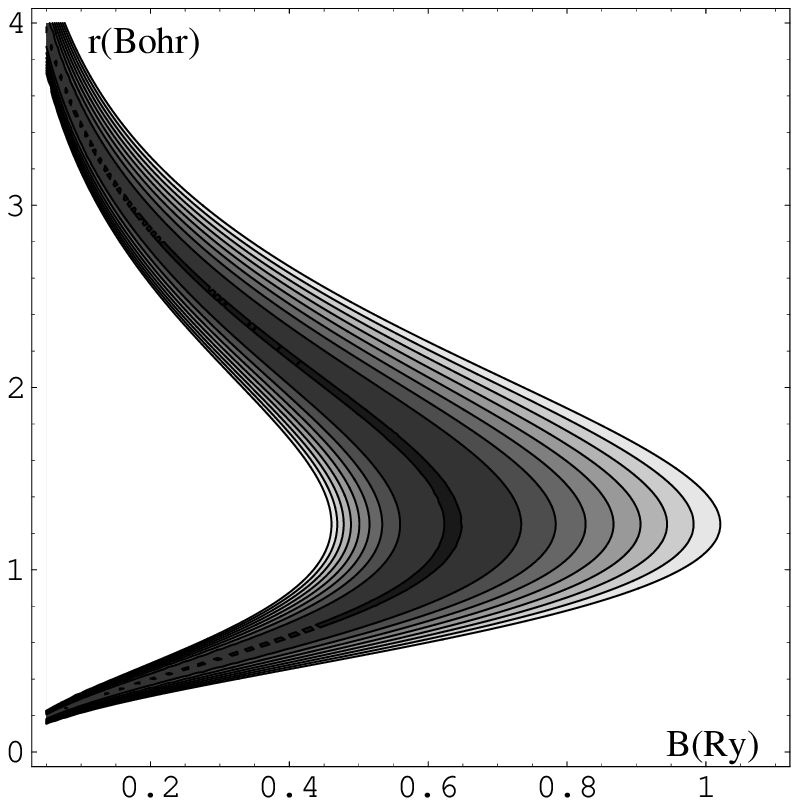}
 \end{minipage}
\caption{Two contour plots of the minimized deviation of the Concurrence as a function of the magnetic field $\mathcal{B} \, (Ry)$ and of the internuclear distance $r$, as given by (\ref{JrHint}). The range of values divided by the contour lines is $[ -0.038,\, 0,04 ] $ for the left panel and $[ -0.03705,\, -0,03000 ] $ for the right one that approximatively corresponding to the black area in the left panel. }
\label{DCurrHy2}
\end{figure}

However, from Figure \ref{DCurrHy2} one can see that the minimized deviation of the
Concurrence takes one minimum for relatively large intensity of
the magnetic field ( say $\mathcal{B} \geq 0.6 \; Ry$), while for weak
fields two minima appear, corresponding to the situation depicted
nearby.

In correspondence of the same values considered above, for $\mathcal{B} =
0.5\; {\rm Ry}$ the function $\Delta {\mathcal C}_{\alpha'_{min}}(r)$ has two minima at $r = 0.79 $ and  $r = 1.88 $, while  for $\mathcal{B} =
0.375\; {\rm Ry}$ they are located at $r = 0.60$ and $r = 2.25$.
So one sees that the resulting \emph{equilibrium configurations} are not
much very close to the experimental one. The equilibrium configuration more closest to the experimental one is the minimum occurring at $r=1.88$ ($\mathcal{B}=\frac{1}{2}$ Ry) for the function $\Delta {\mathcal C}_{\alpha_{min} '}(r)$.

One sees that one of the resulting equilibrium configurations is only roughly close to the experimental one.

In other words, to conclude monitoring numerically $\mathcal{B}$ the equilibrium configuration more closest to experimental one in the minimum occurring at $r= 1.88$ for $\mathcal{B}=\frac{1}{2} \; Ry$ and at $r=2.25$ for $\mathcal{B}=0.375 \; Ry$ for the function $\Delta \mathcal C _{\alpha' _{min}} (r)$.

\section{A quantum chemical framework to compare entanglement and correlation energy \label{qcf}}

In this Section we represent the results produced in \cite{hua05}, where the electron entanglement in the Hydrogen molecule, calculated by the von Neumann entropy of the reduced density matrix $\rho _{1}$, is obtained starting by the excitation coefficients of the wave function expanded by a configuration interaction method:
\begin{displaymath}
S\Big(\rho_{1}^{CISD}\Big)=-Tr\Big(\rho_{1}^{CISD}log_{2}\rho_{1}^{CISD}\Big)=
\end{displaymath}
\begin{displaymath}
=-\Big(\sum_{i}^{m-1}|c_{1}^{2i+1}|^{2}+\sum_{i=1}^{m-1}|c_{1,2}^{2i+1,2i+2}|^{2}\Big)log_{2}\Big(\sum_{i}^{m-1}|c_{1}^{2i+1}|^{2}+\sum_{i=1}^{m-1}|c_{1,2}^{2i+1,2i+2}|^{2}\Big)+
\end{displaymath}
\begin{equation}
-\Big( |c_{0}|^{2}+\sum_{i=1}^{m-1}|c_{2}^{2i+2}|^{2}\Big)log_{2}\Big( |c_{0}|^{2}+\sum_{i=1}^{m-1}|c_{2}^{2i+2}|^{2}\Big),
\end{equation}
 
where $c_{1}$ is the coefficient for a single excitation, and $c_{1,2}$ is the double excitation (in Appendix A of \cite{mio} more details are shown).

\begin{figure}[!h]
\begin{center}
\includegraphics[angle=-90, width=0.55\textwidth]{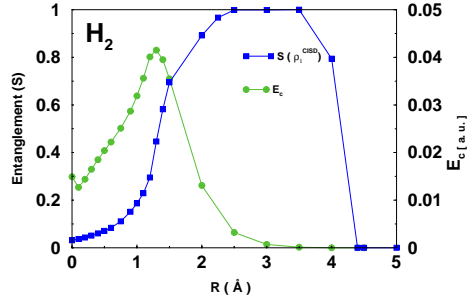}
\caption{Comparison between the entanglement, calculated by the von Neumann
entropy of the reduced density matrix, and the electron correlation energy in the Hydrogen molecule.}
\label{figvecchiomio}
\end{center}
\end{figure}

In this framework, entanglement ($S$) and correlation energy $({\cal E}_{corr})$, as functions of nucleus -- nucleus separation are those in Figure \ref{figvecchiomio}

By the results given by this model, we want to discuss and to suggest some answers to the questions $i)$ and $ii)$ presented in \textbf{Remark 1}. Even if, in order to represent correlation energy and entanglement, we use two different scales, in Figure \ref{figvecchiomio} we can see that entanglement has a small value in the united atom limit after it is growing for small distances till it arrives at a maximum value then it decrease till it assumes zero value at the separated atom limit and it is exactly the progress of the correlation curve.

In order to compare the entropy $S$ with the electron correlation energy ${\cal E}_{corr}$, we rescale $S$ with the parameter $\alpha _{min}$ calculated with some procedure illustrated in Eq. (\ref{deviation}) and Eq. (\ref{-0/69}) replacing the integration variable $\lambda$ with $R$; in this way we extract
\begin{equation}
\alpha =\frac{\int{{\cal E}_{corr}S_{vN}dR}}{\int{S_{vN}dR}}\approx 0.009.
\end{equation}

\begin{figure}[!h]
\begin{center}
\includegraphics[angle=-90, width=0.4\textwidth]{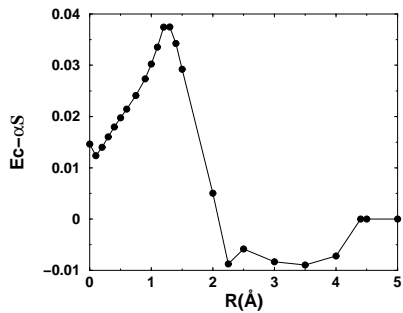}
\caption{$\Delta S_{\alpha _{min}}$ for the $H_{2}$ molecule as a function of nucleus--nucleus distance.}
\label{grDeltaH2}
\end{center}
\end{figure}

The corresponding $\Delta S_{\alpha_{min}}= {\cal E}_{corr}-\alpha S$ allows us to answer to the question $ii)$; in fact, as it is shown in Figure \ref{grDeltaH2}, the vanishing point of $\Delta S_{\alpha _{min}}$ is, according to the two --spin Ising model, nearby $R\approx 2$ $\AA$ that corresponds to the equilibrium configuration of the Hydrogen molecule.

\section{Differences between the Configuration Interaction approach and the two--spin Ising model \label{CI2sIm}}

The model proposed in Sec. \ref{primo} provides us with a measurement of entanglement: indeed, Eq. (\ref{SvNreg1}) describes the von Neumann entropy as a function of coupling constant $\lambda$, for small $\lambda$. By using Eq. (\ref{correlazio}), we can express $\lambda$ in terms of correlation energy and substituting it in  Eq. (\ref{SvNreg1}) we can obtain the variation of $S_{vN}$ in terms of ${\cal E}_{corr}$.

\begin{equation}
S_{vN}=-\frac{\Big\{{\cal E}_{corr} Log\Big(\frac{{\cal E}_{corr}}{2({\cal E}_{corr}+2)} \Big)+({\cal E}_{corr}+4) Log\Big(\frac{{\cal E}_{corr}+4}{2({\cal E}_{corr}+2)} \Big) \Big\}}{({\cal E}_{corr}+2)Log4}. \label{SfunzE}
\end{equation}

In order to calculated the coefficient of proportionality among $S_{vN}$ and ${\cal E}_{corr}$ we make an expansion of $S_{vN}$ for ${\cal E}_{corr} \rightarrow 0$ (or equivalently for $\lambda \rightarrow 0$) at the first order, obtaining a straight line characterized by an angular coefficient given by $m_{S_{vN}({\cal E}_{corr})}=(\frac{1}{4})(1+\frac{1}{Log2})$.
Since this behavior is uncorrect to represent the logatithmic singularity of $S_{vN}$ in the origin, we make an expansion of Eq. (\ref{SfunzE}), preserving the logarithmic deviation, and we obtain an expression of the form 
\begin{equation}
S_{vN}=A {\cal E}_{corr}+ B {\cal E}_{corr} Log ({\cal E}_{corr}), \label{concoeff}
\end{equation}
where  $A=1/2$ and $B=-1/(4Log2)$.
\begin{figure}[!h]
\begin{center}
\includegraphics[angle=0, width=0.4\textwidth]{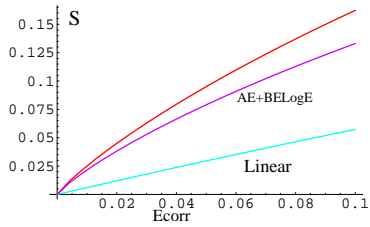}
\caption{A comparison among the behavior of Eq. (\ref{SfunzE}) and its linear approximation and the logarithmic one, for the Ising model.}
\label{leapprox}
\end{center}
\end{figure}

\begin{figure}[!h]
\begin{center}
\includegraphics[angle=-90, width=0.4\textwidth]{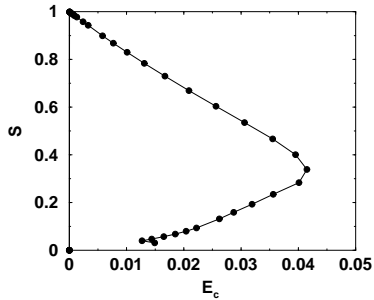}
\caption{A correspondence of ${\cal E}_{corr}$ and  $S_{vN}$ calculated by the  numerical procedure suggested by \cite{hua05}}
\label{grS_Ec.eps}
\end{center}
\end{figure}

In order to compare the behavior of $S_{vN}$ in Eq. (\ref{SfunzE}), we have organized the numerical data, calculated with the method proposed in \cite{hua05}, by making  a correspondence between each value of ${\cal E}_{corr}$ and its respective value of $S_{vN}$, obtaining the plot in Figure \ref{grS_Ec.eps}

\begin{figure}[!h]
\begin{center}
\includegraphics[angle=-90, width=0.4\textwidth]{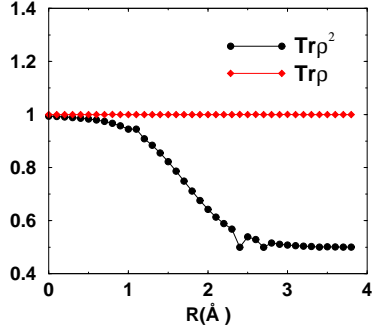}
\caption{The increasing of the degree of mixing in the two electron state: in black we depict the trace of $\rho$, in red the trace of $\rho^{2}$.}
\label{grdendenq.eps}
\end{center}
\end{figure}

Of particular significance is the fact that, in the range where $S$ is monotonically increasing, the correlation energy has its maximum, consequently $S$ seems to be not a function. Moreover, it is important to note that  ${\cal E}_{corr}$ begins to decrease for $R>1$ $\AA$, region where the states become mixed, i. e. ,$Tr \rho \neq Tr \rho^{2}$; as depicted in Figure \ref{grdendenq.eps}.

Probably, for this reason, the procedure adopted in \cite{hua05} seems to be not correct: the density matrix, in fact, is calculated starting by the excitation coefficient of a wave function obtained developping with the Configuration Interaction Single Double method a \emph{pure} two electrons state.

\begin{figure}[!h]
\begin{center}
\includegraphics[angle=0, width=0.4\textwidth]{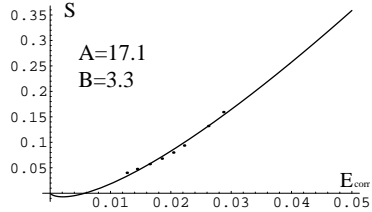}
\caption{A fit of $S_{vN}$ as a function of ${\cal E}_{corr}$, around the origin, with a function of the form $F=A{\cal E}_{corr}+B{\cal E}_{corr}Log({\cal E}_{corr})$ whose coefficients $A$ and $B$ assume the numerical values in Figure. }
\label{show}
\end{center}
\end{figure}

However, even if we consider only the first branch of the plot in Figure \ref{grS_Ec.eps}, i.e. , the numerical values of $S_{vN}$ corresponding with increasing values of ${\cal E}_{corr}$, and we fit the values around ${\cal E}_{corr} \rightarrow 0$ with a $F=A{\cal E}_{corr}+B{\cal E}_{corr}Log({\cal E}_{corr})$ we draw out numerical values of the coefficient different from the ones used in Eq. (\ref{concoeff}). This result is shown in Figure \ref{show}.

In particular the arithmetic sign of the coefficient $B$ in the two models are opposite and this implies the opposite concavity of the curve.

This fact, clearly demonstrates a not satisfactory agreement between the Ising model and the one proposed in \cite{hua05}.

\section{Concluding remarks \label{fine}}

We have explored the role of entanglement in the model of two qubits describing the Hydrogen molecule (\ref{2Ising}), considered as a bipartite system. In our discussion we have limited to the ferromagnetic case governed by the interaction coupling parameter $J>0$.

The concept of entanglement gives a physical meaning to the electron correlation energy in structures of interacting electrons. The entanglement can be measured by using the von Neumann entropy or, alternatively, the notion of Concurrence \cite{woo98}. To compute the entanglement it is convenient to consider two Regions, say \textbf{I} and \textbf{II}, which provide two different reduced density matrices. The entropy  turns out to be an increasing function of the coupling constant $\lambda$ in Region \textbf{I}, but the state under consideration is maximally entangled in Region \textbf{II} indipendently from the anisotropy parameter $g$.

An interesting result is that for large coupling constants the entropy approach 1, meaning that all levels are equiprobably visited by the considered spin.

For weak interactions, at the boundary point $\lambda_{b}=\frac{2}{\sqrt{1-g^{2}}}$ the von Neumann entropy admits a discontinuity, indicating a crossing of the lowest eigenvalues and, in a more general constext, a quantum phase transition \cite{sac01}.

In Sec. \ref{entcorr} a comparison between the entanglement and the correlation energy is performed.

To quantifying the entanglement we resort to the reduced density matrix. The entanglement can also be measured by exploiting the concept of Concurrence.

The entanglement measure is always bounded, while the energy correlation, ${\cal{E}}_{corr}=|{\cal{E}}_{0}|-2=\sqrt{4+\lambda^{2}}-2$, is a divergent function of $\lambda$. This fact tells us that to look for simple relations valid on the whole $\lambda-$axes has no sense.  

Thus, by limiting ourselves to weak couplings, we have minimized the mean square deviation given by Eq. (\ref{deviation}). This procedure leads to the value $\alpha_{min}\approx - 0.691217$ for the minimizing parameter (see Eq. (\ref{-0/69})).

Sec. \ref{primo} contains a comparison between the von Neumann entropy and the Concurrence.

Such a comparison is illustrated in Figure \ref{confronto}, for two spin system as a function of the coupling $\lambda$ for $g=1$.

Some important points are commented in \textbf{Remark 1} and \textbf{Remark 2}  .

In Figure \ref{aroundminS} the deviation $\Delta S_{\alpha}$ and its derivatives with respect to $\lambda$ are computed and $\alpha_{min}$ is evaluated for $\alpha$ ranging in the interval $-1.29 \leq \alpha \leq -0.091$.

In Figure \ref{aroundminC} the minimized Concurrence deviation $\Delta {\mathcal C}_{\alpha ^{(i)}_{\alpha '}}$ for the four eigenstates of the 2-spin model is shown.

We point out the existence of a perfect symmetry among the Concurrence deviations for pairs of eigenstates of opposite eigenvalues.

Formula (\ref{JrHint}), due to Heitler--London \cite{hei27}, is reported, where the position dependent exchange coupling $J(r)$ is expressed in term of the length $r$ of the nucleus--nucleus separation in the Hydrogen molecule.

To conclude, the magnetic field $\mathcal{B}$ has been monitored such that the equilibrium configuration more closest to the experimental one, $r \approx 2.00$, is the minimum occurring at $r=1.88$ for $\mathcal{B}=\frac{1}{2} \; Ry$ and $r=2.25$ for $\mathcal{B}=0.375 \; Ry$ for the function $\Delta {\mathcal C}_{\alpha '_{min}}(r)$. 

We observe also that in the intermediate range of values, i. e., for $0.6 \leq \lambda \leq 1$, the two functions $SvN$ and the correlation energy are almost proportional within the $10 \%$.

However, when we organized the pairs of points (${\cal E}_{corr}, S_{vN}$) calculated by following the procedure described by \cite{hua05}, it is clear that the von Neumann entropy cannot be considered a function of correlation energy. The principle cause is that the function ${\cal E}_{corr}$ presents a maximum in the region where $S_{vN}$ is monotonically increasing.

The reversing behavior of correlation energy occurs in correspondence with an increase of the mixing degree of the two electrons state. The function ${\cal E}_{corr}$ in terms of the nucleus -- nucleus distance $R$, increases till the state is pure, on the contrary, when $Tr(\rho^{2})$ becomes discordant from $Tr(\rho)$, the function ${\cal E}_{corr}$ decreases.

This fact suggests us that the numerical model based on the calculation of $S_{vN}$ starting by the excitation coefficients $c_{i}$, isn't completley correct because the density matrix is obtained as a product of two electron pure states.  
However, even if we consider only a branch of the plot in Figure \ref{grS_Ec.eps}, the function obtained by the two spin Ising model, i. e., Eq. (\ref{SfunzE}), is unsuitable for fitting these numerical data. 

On the basis of our results, essentially grounded on numerical considerations, in the near feature we would  explore more complicated systems of molecules, such as for example the ethylene or other hydrocarbons, and compare these studies with the goals obtained for the Hydrogen molecule. 

\section*{Acknowledgments}
The authors acknowledge the Italian Ministry of Scientific
Researches (MIUR) for  partial support of the present work under
the project SINTESI 2004/06 and the INFN for  partial support
under the project Iniziativa Specifica LE41.

\end{document}